# On the uniaxial static stress relaxation in fused silica at room temperature

SAOOD IBNI NAZIR, CHRISTOS E. ATHANASIOU AND YVES BELLOUARD

*Galatea Lab, STI/IMT, Ecole polytechnique fédérale de Lausanne (EPFL), Rue de la Maladière, 71b, Neuchâtel CH-2002, Switzerland*
*\* Corresponding author: saood.nazir@epfl.ch*

The question of whether static stress in fused silica relaxes at room temperature is still under debate. Here, we report experimental data investigating stress relaxation dynamics in fused silica at room temperature and up to 2 GPa stress levels under static loading conditions. Our measurements are performed using an innovative combination of methods; a femtosecond laser is used to accurately load a monolithic microscale test beam in a non-contact manner, fabricated using femtosecond laser processing and chemical etching, while the stress is continuously monitored using photoelasticity over an extended period spanning several months, in both dry and normal atmospheric conditions. The results are of practical importance, not only for photonics devices making use of stress-induced birefringence but also for flexures subjected to static loads.

## 1  Introduction

Thanks to its interesting mechanical, thermal, chemical, dielectric, and optical properties, fused silica is of technological importance and extensively used in diverse applications covering optical, electronics, biological, and chemical engineering. Recently, emerging three-dimensional micro-manufacturing processes, such as femtosecond laser exposure combined with wet chemical etching, have triggered a special interest for fused silica as a key material in a variety of micro-technologies, such as lab-on-chip [1]–[3], photonic devices [4], [5], micro-actuators [6], micro-mechanical devices [7], [8], and also for packaging applications [9].

Some of these small-scale devices experience high static stress, sometimes exceeding GPa levels, either due to the presence of laser-modified zones in the bulk of the material or while in operation for deformable devices. The long-term behavior of such devices remains to be investigated. More generally, the question of whether static stress in fused silica relaxes at room temperature, in the presence of moisture or not, is still under debate.

Fracture and stress relaxation in silica is a multifaceted problem, difficult to study experimentally, in particular at the microscales [10], [11]. Surface flaws such as micro-cracks that may result from the manufacturing process [12] can act as stress concentration points, leading to a reduction in fracture strength [13], [14]. Under the application of a static stress and in the presence of moisture, such cracks can nucleate and result in delayed failure [15]–[17]. A minuscule amount of water can significantly impact the mechanical properties such as lowering of Young's modulus, density, viscosity, and glass transition temperature [18], [19]. The presence of stress also affects water solubility as well as its diffusivity in the material [20]. Due to the volume changes that occur during a glass-water reaction [19], [21], the reaction is stress-sensitive, and stress can act as a catalysis for the chemical reaction. At low temperature (< 250℃), the presence of *compressive* stress increases water solubility while reducing the diffusion coefficient. At higher temperatures (> 650℃), opposite behavior is observed. Interestingly, both trends are reversed in the presence of *tensile* stress. These interfacial mechanisms are complex to investigate and to date not fully understood at room temperature, in part due to the lack of comprehensive reported data.

In the sequel, we explore how fused silica behaves when subjected to *high static tensile stresses* (up to 2 GPa) at *room temperature*, *with* or *without moisture*, and over an *extended period* (spanning several months). To this end, we use a dedicated monolithic micro-tensile tester device fabricated with an advanced manufacturing process based on the combination of femtosecond laser exposure and wet chemical etching. The micro-tensile tester allows for not only loading a micro-scale beam in a pure uniaxial tensile mode to precise and controllable stress levels but also for maintaining the tension throughout the measurement duration without further intervention and with only one material involved. Stress measurements are facilitated through photoelasticity by measuring the retardance of a laser beam propagating through the stressed region. The stressed state is attained inside a well-controlled atmosphere of dry $N_2$ gas, while subsequent retardance measurements are done in both dry and ambient humid conditions.

## 2  Monolithic miniature tensile tester – design and working principle

The monolithic device, shown in Fig. 1 is an optimized version from our previous work on micro-tensile testers [22]–[24]. The device is fabricated using a process combining femtosecond laser exposure and wet chemical etching (described in detail in [25] and explained briefly later) and integrates both the test specimen and the means to load it precisely under uniaxial tension. To introduce a given stress level in the test beam, we also use femtosecond laser exposure, this time to induce a localized volume expansion in strategic locations within the device. Thanks to a flexure-based linear guidance and interface kinematics, the laser-induced volume changes are channeled to the microscale beam such that, the displacement field is perfectly aligned with the beam's long axis, leading to a pure uniaxial tensile stress. By varying the density of modifications, and the dose of laser exposure, precise stress levels can be achieved in a controlled manner. Using this method, we are able to achieve stress levels exceeding 2 GPa in a few micron-wide beam.

The tensile tester (including the loading region and the test specimen) is small enough to be fitted in an enclosure where the atmosphere is accurately controlled, not only during the stress-loading process performed in a dry atmosphere of $N_2$ gas but also, thereafter, over an extensive period allowing for continuously monitoring the stress evolution through photoelasticity in controlled atmospheres. For measuring the stress in the test beam, we rely on the stress-induced birefringence that induces retardance in the optical path of a light beam passing through the micro-tensile test beam.

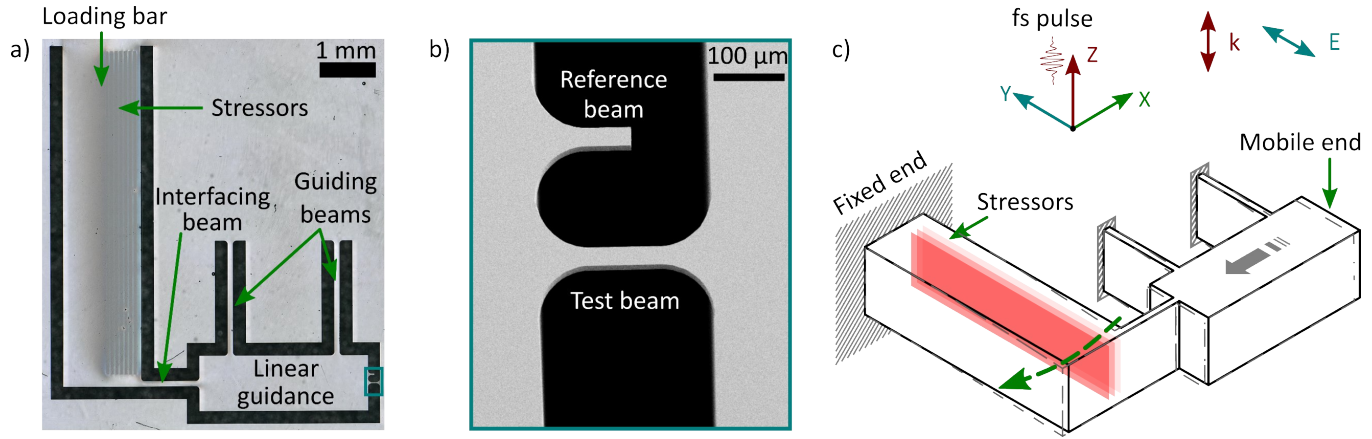

Fig. 1. a) An optical image of the loading device fabricated using femtosecond direct laser writing and wet chemical etching. After etching, femtosecond laser exposure within the bulk of the loading bar results in a curved trajectory of the free end, which is coupled into a linear guidance and channeled into a uniaxial strain applied to the test beam. b) A high magnification image of the test beam and the reference beam (boxed region in (a)). c) An illustrative diagram of the loading mechanism showing the deformed state (in-dash) underneath. The modifications are written along the length of the loading bar with the laser polarization oriented along the writing direction.

The working principle of the device is detailed in Fig. 1. It consists of a loading region (left part of Fig. 1a), where laser affected zones (referred to as stressors from here on) are inscribed in the volume of the device below the surface. Under a certain set of laser parameters, femtosecond laser exposure of fused silica results in nano-porous self-organized structures [26], [27] within the modified zone. These structures exhibit a net volume expansion [28] that can be controlled locally in magnitude as well as in direction [29], [30]. In the particular arrangement considered here, with an expanding material on one side and pristine material on the other side, the loading bar behaves as a bimorph element, resulting in a bending movement (in-plane) and a deflection at its free end. Coupled to this end, an elongated bar pulls on a linear guidance flexure mechanism. The purpose of the elongated beam is to interface the curved trajectory of the bimorph element endpoint with the linear motion of the guidance. The linear guidance consists of two parallel bars and is equivalent to a four-bar parallelogram kinematics in mechanism theory. As the displacements involved here are very small compared to the dimensions of the parallel bars and do not exceed a few microns, the parasitic (out-of-plane) motion in the Z-direction is negligible. The test beam at the end of the guidance connects the guidance to a static frame. The beam is positioned symmetrically with respect to the interfacing bar connecting the loading mechanism to the guidance. This ensures a zero bending moment about the fixed end of the beam and a pure tensile loading mode. For retardance calibration purposes, next to each test beam, a half beam is fabricated and attached to the guidance on one end while the other end extends roughly to the midpoint of the main beam. Being free on one end, this element does not see any stress and is used as a marker for retardance measurements to focus the probe laser at the same point in each measurement. The overall design is compact (7×6 mm$^2$), monolithic, and fabricated in a single step using femtosecond direct laser writing.

To operate the mechanism, stressors are inscribed near the sidewall of the loading bar. Depending on where the stressors are placed within the loading bar, it can bend two ways, towards or away from the linear guidance resulting in compressive or tensile stress respectively in the test beam. In this experiment, however, exposures are only carried out on one side, resulting in a purely uniaxial tensile stress state.

The geometry of the mechanism is optimized to minimize the loading time. Given the marginally small volume expansion, typically 0.01-0.05 % of the modified volume [28], and our limited experimental setup, it would take up to ten hours to reach stresses above 1 GPa. However, in this design, the length of the loading bar effectively amplifies the strain manifold and reduces the exposure time to less than thirty minutes per stressor. This translates to about four hours for reaching a level of 2 GPa. To prevent the formation of stress concentration points at sharp boundaries, a dog-bone shape is adopted for the design of the test beam. The rounding edges have a radius of 50 μm.

Finally, a key point of this device is that it is entirely transparent and the stress state can be measured remotely in an optical manner as will be described later. Due to the small size of a single tensile tester, multiple testers can be fabricated on the same substrate.

## 3 Experimental setup

### 3.1 Tensile tester - fabrication and re-exposure

The devices are fabricated out of a 25 mm × 25 mm × 1 mm silica substrate (synthetic fused silica with low OH content – Corning 7980 0F) using a femtosecond laser fiber amplifier (Yuzu, Amplitude Systèmes, France) delivering 270 fs pulses at 1030 nm. The laser is focused using a 0.4 numerical aperture objective (LMU Microspot, Thorlabs, USA) and the sample is translated beneath the focus with the help of precise positioning stages (Micos, Ultra-HR, Germany). After machining, the samples are slowly etched in a 1.25 % solution of hydrofluoric acid (HF) for about 40 hours. A slow etching step is used to prevent an aggressive attack on the modified regions and to improve the surface roughness as etching is known to affect fracture strength [31]. This is particularly critical for the surface quality of the test beams and to prevent crack nucleation while loading. On each substrate, six micro-tensile testers are fabricated.

After etching, the tensile-testers are re-exposed under the same laser to load the test beams. For this step, the samples are placed inside a sealed miniature environmental chamber with optical access viewports at the top and bottom for laser exposure and subsequent retardance measurements. A continuous flow of dry $N_2$ gas creates a dry atmosphere inside the chamber.

The stressors are written in the form of discrete blocks, each 50 μm-wide in the XY plane, extending from the bottom surface to the top surface of the loading bar (along the YZ plane), and spanning its entire length. To prevent surface ablation, the stressors are embedded inside the volume of the loading bar at a depth of 20 μm from the top and bottom surfaces. Within each stressor, the lateral spacing between two individual lines is set at 2 μm and the planes are spaced by 10 μm along the vertical direction (Z-axis). Near the edges, the stressors are placed 100 μm away to prevent stress concentration and possible crack formation. Between two individual stressors, a wide gap of 50 μm is left unmodified. In this step, a pulse energy of 220 nJ, pulse repetition rate of 500 kHz, and optimal deposited energy of 12 J/mm$^2$ is used. The choice of deposited energy is dictated by an optimization process that achieves the highest exposure to beam deflection efficiency while preventing cracks in the laser-exposed zones [32]. In our case, this translates to a writing speed of 5.9 mm/sec.

During the loading process, the laser power is continuously modulated along the thickness of the loading bar to compensate for parasitic out-of-plane bending. Such bending is caused by non-uniform laser exposure conditions due to spherical aberrations along the thickness of the substrate [22]. Specifically, we modulate the power by 25 % across a thickness of about 1 mm to compensate for power loss near the bottom surface and to create a uniform strain across the thickness of the loading bar. Furthermore, to maintain a uniform deposited energy along the length of the stressors, a buffer is added before the beginning and at the end of each line to account for the acceleration/deceleration phases of the translation stages. This buffer gives the stages enough time to attain a constant speed before arriving at the endpoints of each line.

By varying the number of stressors, the overall strain and the resulting stress in the test beams can be controlled precisely. A calibration curve showing the stress as a function of the number of stressors is shown in Fig. 2. The insets show images of the stressed test beam under bright field white-light illumination and a cross-polarized configuration. In such a configuration, regions under stress allow light to pass through and appear white, while light illuminating the surrounding areas is blocked by the analyzer and appear dark. This calibration is important to predict the stresses beforehand and to cover a broad range of stress levels. The loading level varies nearly linearly as a function of the number of stressors.

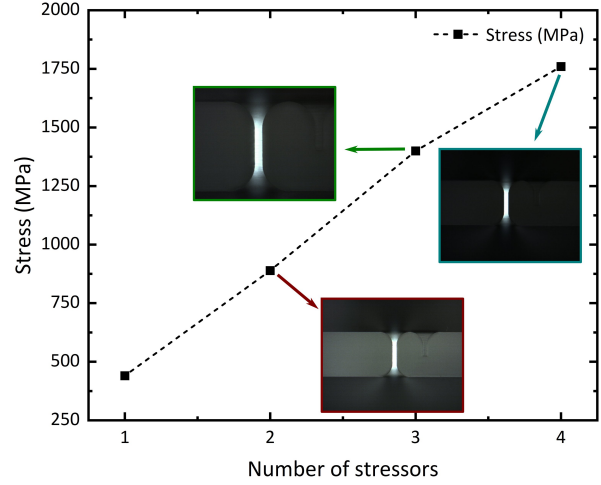

Fig. 2. Stress calibration curve as a function of the number of stressors. A single stressor is 50 μm-wide, 950 μm-thick, and 5.8 mm-long. Photoelastic images corresponding to each stress level are shown in the inset.

*3.2 Stress monitoring*

The stress within the beams is measured using a non-contact method based on photo-elasticity. Although fused silica is isotropic, it can become birefringent when subjected to external stress. In such a case, the principal optical axis is defined by the direction of applied stress. In our measurements, we observe the retardance of a monochromatic laser beam passing through the stressed region. To do this, we use a highly stable and collimated Helium-Neon laser ($\lambda = 633$ nm), which is focused inside the specimens (test beam region) placed in a cross-polarized configuration as shown in Fig. 3. The stress within the test beams is then calculated according to the following equation:

$$\sigma_1 - \sigma_2 = \frac{R}{T(C_1 - C_2)}$$

Where R is the measured retardance, T is the thickness of the test beam, $C_1-C_2 = 3.55\times10^{-12}$ Pa$^{-1}$ is the stress optic coefficient of fused silica, and $\sigma_1$, $\sigma_2$ denote the stresses along the axial and transverse directions respectively. In our case, we consider $\sigma_2$ negligible due to pure uniaxial loading conditions.

The laser output is expanded via a telescope to match the entrance aperture of the focusing objective. The beam is then split using a polarizing beam splitter (PBS) and part of it is incident on a photodiode to monitor power fluctuations of the source. The other half is focused through an optical window using a long-range objective (Mitutoyo 20X, Japan) with a working distance of 20 mm and a numerical aperture of 0.4. The waist at the focus is approximately 2.5 μm and the Rayleigh range is nearly 30 μm (hence larger than the thickness of the measured beams). The choice of a large working distance objective is necessary to access the beams placed inside the environmental chamber. The chamber (see inset of Fig. 3) is mounted on an XY stage to access the four sample slots. Within each slot, a substrate containing six micro-tensile testers can be fitted.

A half-wave plate orients the polarization of the beam at 45° with respect to the principal stress axis in the test beams. After focusing, the beam is recollimated using a telecentric lens and made incident on a variable retarder (Soleil Babinet Compensator, SBC-VIS, Thorlabs, USA) which is used to quantitatively measure the induced retardance in the optical path. The compensator plate inside the SBC can be set to a variable retardance or be used to measure an unknown retardance. Further downstream from the SBC, an analyzer is placed in a cross-polarized configuration with respect to the polarization axis defined by the half-wave plate. In this way, the test beams are placed in a cross-polarization configuration. The output is then focused on an amplified detector to measure the resultant birefringent signal. To precisely focus the He-Ne laser at the center of the beams, an LED ($\lambda = 455$ nm) is used for imaging. The LED is focused using the same objective and the reflected light is focused on a camera using a tube lens. To prevent the back-reflected laser beam from saturating the camera, a 45° angle of incidence (AOI) mirror is inserted before the camera. The mirror is reflective at 633 nm and transmissive at 455 nm. Similarly, to block the unpolarized LED from contributing to the detector signal, a long-pass filter ($\lambda_{cut-off} = 600$ nm ) is inserted after the SBC.

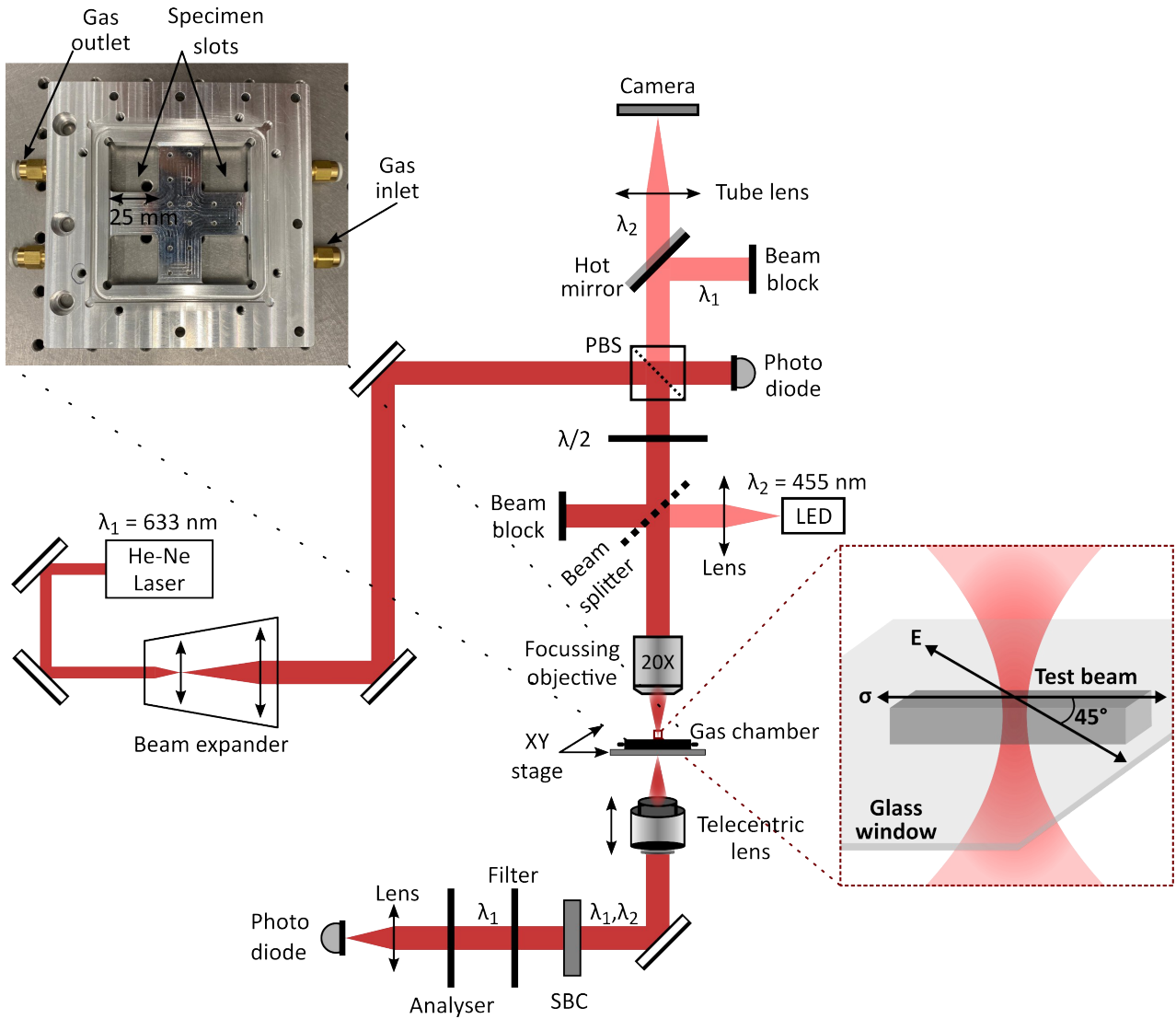

Fig. 3. Schematic of the stress measurement experimental setup. A laser beam from a He-Ne source is focused inside the stressed beams oriented at 45° with respect to the polarization state. The recollimated beam is passed through a second polarizer oriented in a cross-polarized configuration with respect to the incident polarization and focused onto a photodiode. For retardance measurements, a Soleil Babinet Compensator (SBC) is inserted in a zero retardance configuration. To image the beams, an LED is illuminated from the top.

## 4 Experimental results

### 4.1 Stress measurements over time

The beams are loaded sequentially each time with an increasing stress level. After each loading step, the optical retardance through the test beam is measured. A loading step consists of writing a certain number of stressors (depending on the desired stress level) inside the loading region of the tensile tester where each stressor consists of nearly 2500 laser written lines. This sequence is repeated until all the beams (six) on a single substrate are loaded. To minimize alignment errors during positioning of the chamber on the measurement setup, two precision alignment pins are used.

For our measurements, we fabricate two sets of samples. In the first set, the thickness (along Z) of the test beams is relatively large (ranging between 15.9 – 26.4 µm) and varies substantially between the individual beams. This set contains a total of seven loaded beams. The second set with four beams has a much lower beam thickness (ranging between 5.3 – 8.7 µm). The thickness of each tested beam is measured precisely using a confocal microscope (Keyence VK-X1000 Series, Japan). The accuracy of this measurement is estimated to be within 100 nm. The beam dimensions are given in Table 1.

Table 1. Beam dimensions of the loaded specimens measured with a confocal microscope. With the objective used (50X), the measurement errors are within 0.1 µm along both axes. All dimensions are in µm.

| Test beam | Width | Thickness |
|---|---|---|
| A1 | 22.1 | 15.9 |
| A2 | 21.8 | 13.2 |
| A3 | 22.3 | 15.9 |
| A4 | 22.7 | 19.1 |
| A5 | 27.4 | 24.5 |
| A6 | 28.6 | 25.3 |
| A7 | 26.9 | 26.4 |
| B1 | 30.3 | 5.3 |
| B2 | 25.0 | 8.1 |
| B3 | 24.2 | 8.7 |
| B4 | 26.0 | 8.1 |

In the first set, the beams are loaded successively one after another, and the stress is measured after each loading step, while in the second set, each beam is loaded first and the stress is monitored for a few days before loading the next beam. The stress measurement results are shown in Fig. 4. To make the plots easier to understand, we plot the data of each set separately.

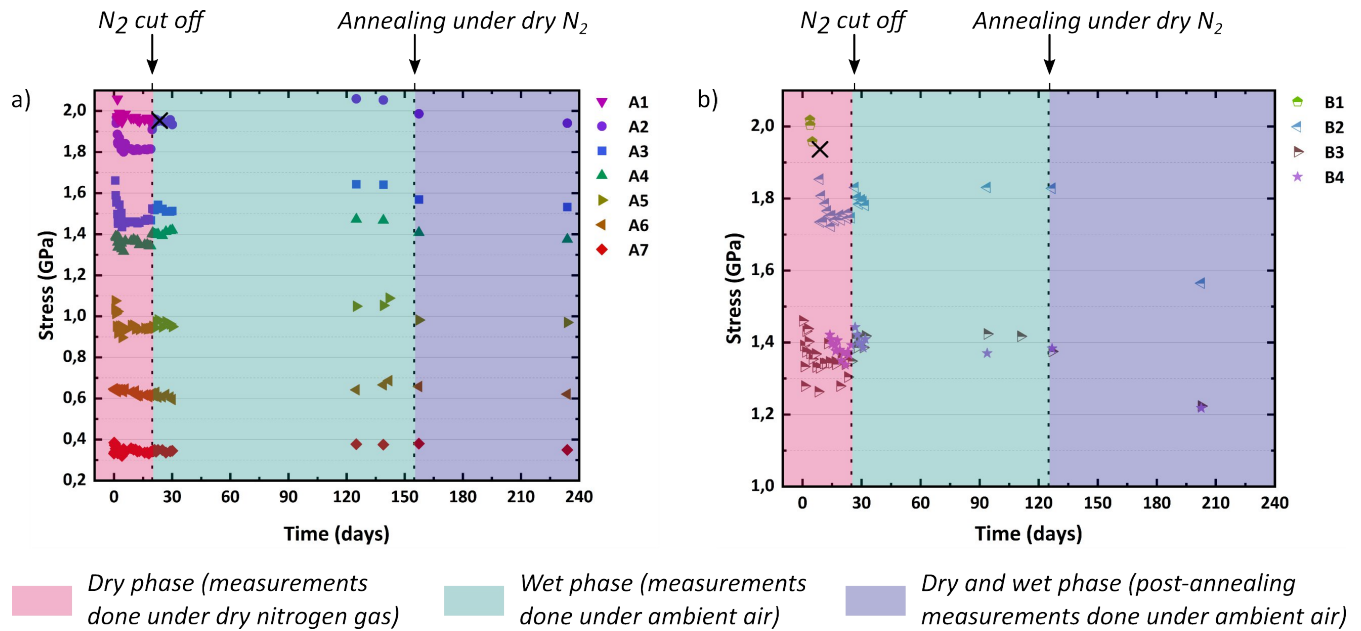

Fig. 4. a,b) Stress aging measurements on fused silica at room temperature and under dry and moist conditions. In the plots, the first dotted line indicates the stopping of dry $N_2$ gas flow and exposure to ambient atmosphere (T = 22 ± 1°C, RH = 50 ± 15 %). The second dotted line and the data point after mark the measurement carried out after the samples were annealed at a low temperature of 300°C for 30 hours. For specimens A1 and B1, the cross denotes failure of the loading bar (more information in supplementary section).

After loading, the retardance is monitored for three weeks while the samples remain inside a dry atmosphere of nitrogen gas. The $N_2$ gas is 99.99 % pure and contains very low concentrations of $O_2$ (< 50 ppm) and $H_2O$ (< 30 ppm). During this first period (referred to as the 'dry phase' in Fig. 4), a comparatively fast decay (up to 20 MPa/day in some cases) is observed immediately after loading followed by a relatively stable stress state. This decay is particularly prominent in beams loaded above 1 GPa stress. After three weeks (marked by the first dotted line in Fig. 4), the $N_2$ flow is stopped and the samples are exposed to normal atmospheric conditions (referred to as 'wet phase' in Fig. 4) as observed in our laboratory. There, the average measured temperature is 22°C with a fluctuation of ±1°C, while the average humidity is 50 % with a fluctuation of ±15 %. As the samples are exposed to these conditions, a slight increase in stress is observed almost immediately (the next measurement is recorded nearly 17 hours after letting the ambient atmosphere flow inside the chamber). This increase in stress is independent of the number of laser exposed patterns but rather appears to depend on the thickness (and hence, the overall laser-machined area) of the test beams as shown in Fig. 5. After this point, further aging measurements are carried out under normal atmospheric conditions.

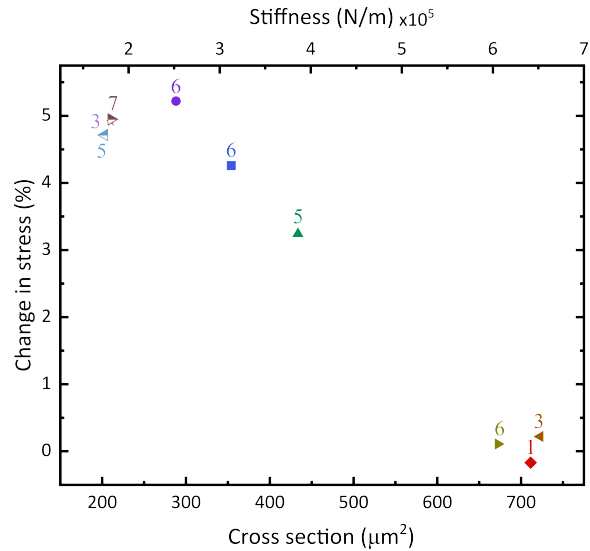

Fig. 5. Change in stress plotted against the cross-sectional area and stiffness of the test beams. The numbers beside each data point indicate the number of laser-written stressors.

To understand this increase in stress on exposure to moist conditions, we leave the beams exposed to these conditions and observe them again after a prolonged period spanning almost 12 weeks. During this prolonged exposure to moist conditions, and like before, a further increase in stress is observed. At this point (indicated by the second dotted line in Fig. 4), the samples are annealed at a low temperature of 300℃ for 30 hours and under a dry atmosphere of $N_2$ gas. The choice of this temperature is dictated by the fact that high-temperature annealing (above 600℃) is known to permanently change the volume expansion of laser-modified patterns [33], thus altering the loading conditions. At 300℃, no such effect is observed, yet the temperature is sufficiently high to cause desorption on the surface. After annealing, the samples are cooled down slowly at a rate of 1℃/min to avoid the buildup of additional stress due to fast quenching.

After annealing, the samples are put back in the chamber and under a constant flow of dry $N_2$ gas. During the transfer from the furnace to the environmental chamber, the samples are briefly exposed to the ambient atmosphere. As the transfer time is short (< 10 min, and thus negligible in comparison to the timescale of the dynamics reported above), we do not expect further adsorption to occur. Further measurements reveal a decline in stress in nearly all of the stressed beams. To understand the increase in stress upon exposure to a humid environment and the decay after annealing (while under a dry atmosphere), we re-expose the beams to humid conditions by cutting off the flow of nitrogen and letting atmospheric conditions fill the chamber. Since the stress levels showed an increase on earlier exposure to ambient conditions, we expect a similar trend. The beams are exposed for a longer duration than earlier before additional measurements are carried out. However, the measured stress shows an opposite behavior. Unlike earlier, we observe a decay in stress. This is particularly substantial for the beams B2, B3, and B4. In Fig. 4, this part of the measurement is referred to as 'dry and wet phase'.

## 5  Discussion and interpretation

At low temperature (< 250℃), the reaction of water with silica is believed to be governed by the following equation:

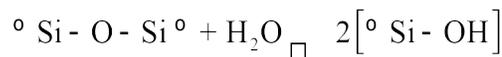

The rate of this reaction is influenced by temperature, vapor pressure, and stress (the stress can be either residual or externally applied). Even though our measurements are carried out at room temperature conditions, the presence of high stress can accelerate the diffusion process. Roberts and Moulson [34] have also suggested that the reaction in Eq. 2 may take place at a detectable rate at room temperature if activated by sufficiently high stress. Furthermore, water diffuses differently depending on whether it is present in vapor form or liquid state. In liquid form, its concentration in the glass increases rapidly reaching a saturation value. In vapor form, the concentration does not achieve a constant value but builds up over time eventually reaching a steady-state value.

The fabrication of our samples involves a two-step process, femtosecond direct laser writing followed by wet chemical etching in a solution of 1.25 % (by volume) hydrofluoric acid. During etching, as there is prolonged contact with liquid water (~ 40 hours), it is plausible to expect a higher water content in the samples after this etching step. At the beginning of our experiment (dry phase), a rather fast relaxation is observed at stresses exceeding 1 GPa. As there is nearly no water present within the chamber, water present within the samples may promote stress relaxation. Fused silica is known to be permeable to helium [35] and hydrogen [36], however, the atomic size of nitrogen makes its penetration impossible. Therefore, the role of nitrogen in relaxation can be ruled out.

In the micro-tensile tester, the primary stress is tensile in nature and occurs within the test beams. However, as shown in Fig. 6, moderate compressive stress is also present within the loading bar in the region surrounding the stressors (region $S_c$). There, as the modified region undergoes expansion, compressive stress builds up in the surrounding pristine material. Similar compressive stress is also present in the flexure beams of the linear guidance, however, its magnitude is negligible in comparison. The presence of compressive stress in the pristine region of the loading bar may aid water solubility on that side (during the wet phase). In comparison, the tensile stress present on the modified side (region $S_t$) lowers water solubility. As a result, Young's modulus of the pristine region ($S_c$) reduces leading to a reduction in its stiffness. The adsorption of water molecules causes the replacement of some bridging oxygen atoms, which are replaced by an Si-OH configuration. This leads to a reduction in tensile strength of the pristine region, thus promoting further bending of the loading bar. Since this bending occurs away from the micro-tensile beams, an additional tensile force acts on the beams leading to an increase in the measured stress. As water is present in vapor form, its concentration within the loading bar is expected to grow over time, furthering the stress increase according to this mechanism. This is corroborated by the increase observed during the prolonged exposure to such conditions (the region between the dotted lines in Fig. 4). Another event in support of this hypothesis is the failure observed in the A1 specimen. Although the highest magnitude of stress is present within the test beam, yet, on exposure to humid conditions, failure occurs within the loading bar instead. This suggests that events occurring within the region under compression are dominant over those occurring within the stressed beams.

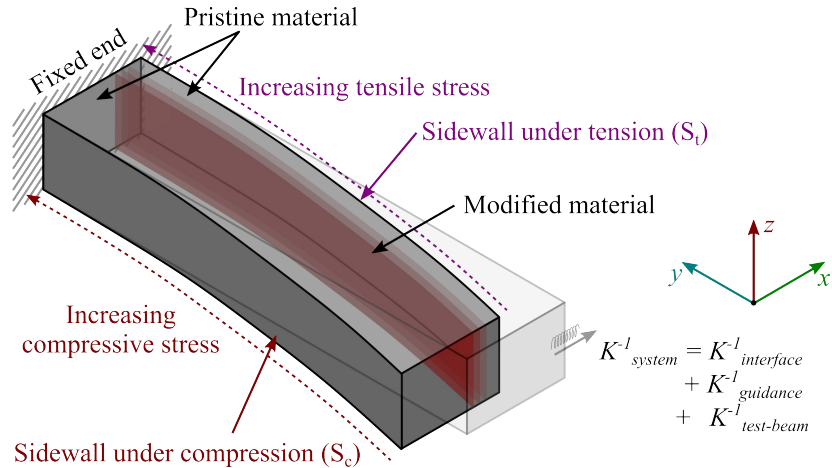

Fig. 6. A schematic of the loading bar showing different regions under compressive and tensile stress. The bimorph-like loading bar consists of three main regions – a pristine region under compression ($S_c$), stressors, and a pristine region under tension ($S_t$). At its free end, the stiffness of the system ($K_{system}$) acts against the bending direction.

Water penetration in glass is known to increase its volume [19], [21]. If the expansion is constrained to occur, such as on the surface, compressive stresses develop. This further enhances water solubility (at low temperature) leading to additional compressive stresses. In such a case, the resultant outcome would also be an increase in stress in the test beams. Overall, a weakening of the pristine structure surrounding the stressors or the generation of a compressive surface stress will both promote further bending of the loading bar and lead to an increase in measured stress as observed in our experiments (at the start of the wet phase). In either case, such an increase should vary inversely with the cross-sectional area (or the stiffness) of the micro-tensile beams as corroborated by the data in Fig. 5.

At high temperatures (> 650℃) water mostly exists as silanol groups in silica, which are responsible for its volume expansion. At low temperatures (< 250℃) water exists both as molecular water as well as in the form of hydroxide.

Building upon our hypothesis that water adsorption could be a possible cause for the increase in stress, it is only logical to reverse the process through desorption. To do so, we choose a moderate annealing temperature of 300℃. Ideally, annealing at an elevated temperature would enhance the desorption of molecular water. This would however result in an ambiguity in our measurements due to the change in volume expansion of the stressors [37]. The removal of molecular water, particularly from the pristine part of the loading bar would, through a similar explanation, cause a decrease in the measured stress.

Post the annealing step, and when the beams are re-exposed to a humid environment (dry and wet phase), there is a rather small decrease in stress in nearly all of the loaded specimens. Of interest, however, are specimens B2, B3, and B4, which show a much larger decrease in the measured stress. The specimens B3 and B4 decay nearly by the same amount (11.1 % and 11.9 % respectively), whereas B2, with higher initial stress, decays more (14.4 %). On further inspection, these relaxed specimens reveal formation of point-like defects within the stressed beams. These defects appear on the bottom surface of the beams, which is laser-exposed during the fabrication step, and thus has greater surface roughness than the pristine top surface. Furthermore, as shown in Fig. 7, the defects appear to be concentrated near the central part of the beam (with high stress) and almost disappear near the low-stress ends. It is though unclear how they are responsible for stress relaxation.

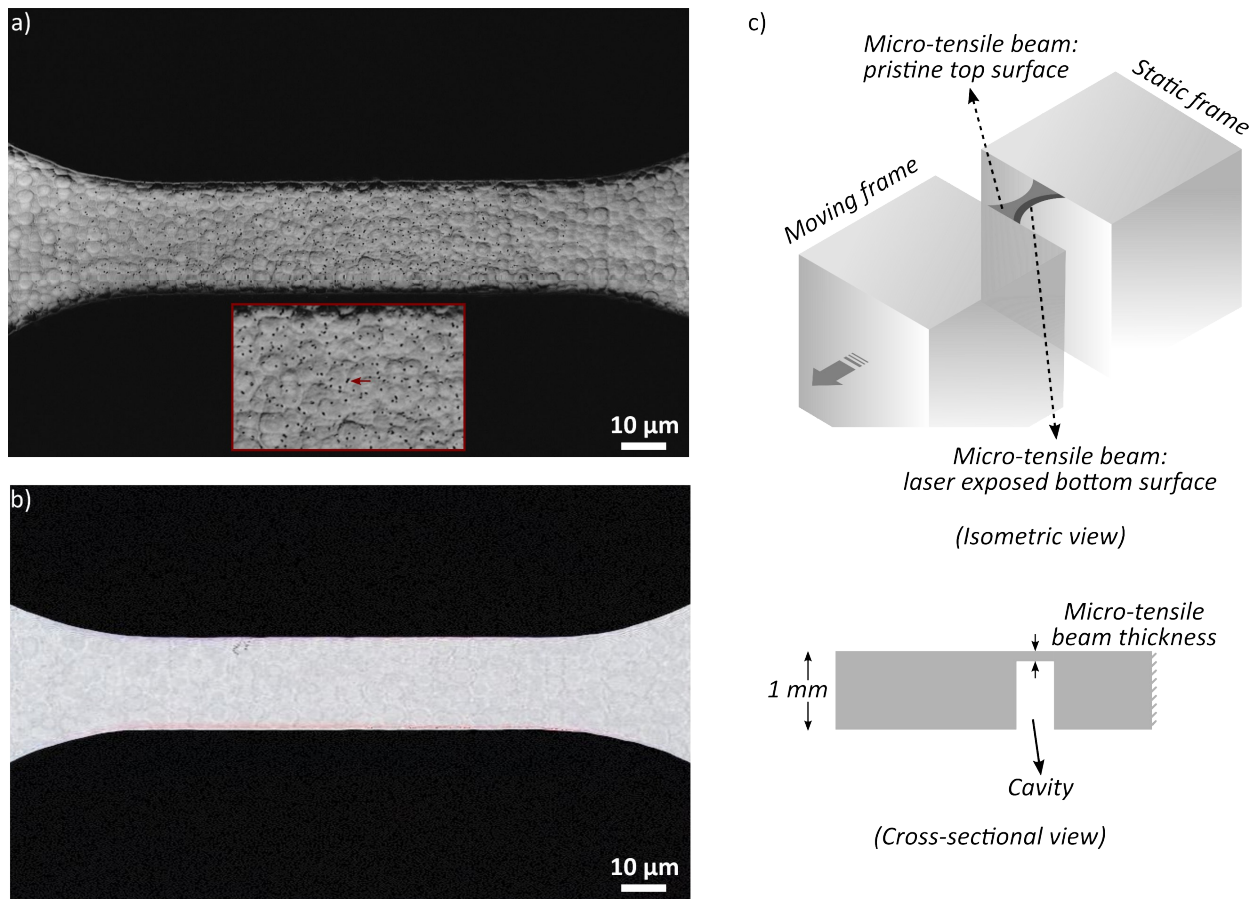

Fig. 7. a) An optical image showing point-like defects observed on the laser-machined bottom surface of one of the beams with highest stress relaxation. The defects, shown in the inset, are concentrated in the middle part of the beam with high stress and disappear towards the low-stress ends. b) Optical image of the pristine top surface of the same beam. c) A 3D illustration highlighting the micro-tensile beam position within the tensile tester. On the bottom surface of the beam, material is removed during fabrication resulting in a cavity.

To validate our hypothesis that water intake during the etching step could be responsible for the relaxation observed initially, we fabricate a new sample with two tensile testers. After etching, this sample is annealed at 800℃ for 300 hours to remove any water that might have penetrated the sample during this etching step of the fabrication process. After annealing, the sample is placed within the chamber with a flow of dry $N_2$ gas and loaded as before. The stress measurement results are shown in Fig. 8.

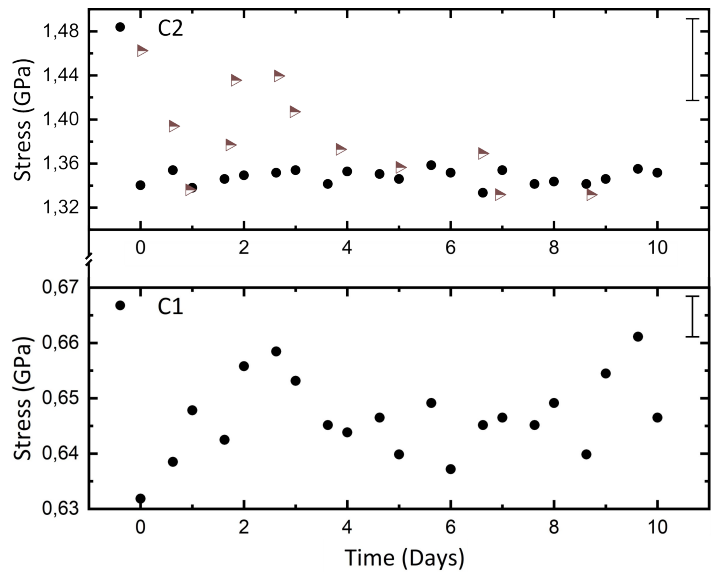

Fig. 8. Stress versus time measurements of an annealed sample. Post etching (1.25 % HF) and before stress loading using femtosecond laser exposure, the sample is annealed at 800℃ for 300 hours to remove any water that might penetrate the sample during the 40-hour etching process. The measurements show a stable stress state unlike the fast initial decay observed for non-annealed samples. For comparison, the data for sample B3 is replotted (brown triangles), which has a similar initial stress as sample C2. The error bars are analytical and do not include other sources of error such as uncertainties in the angle between σ and E, variation in measurement position over time, etc.

Unlike the fast decay observed earlier, a rather constant stress level is seen here. The specimens C1 and C2 have a beam thickness of 7.3 μm and 8.5 μm respectively. These results seem to validate our earlier hypothesis that fast initial decay occurs due to water adsorption during the microfabrication process.

## 6  Conclusion

Using a novel experimental method based on advanced manufacturing of a monolithic tensile tester device, we have studied the behaviour of fused silica beams subjected to high uniaxial tensile stress at room temperature, and in both dry and humid conditions. Our measurements reveal that static tensile stress up to 1 GPa applied to the micro test-beam is stable in magnitude under a dry atmosphere as measured over a period exceeding a year. Above this critical level, and measured up to 2 GPa, we notice an initial relaxation of a few percent while maintaining the structural integrity of the test beams. The mechanism responsible for this observed relaxation is related to the instrument design, where a beam acting as the actuator is loaded in bending. We further show that this relaxation also appears to be directly related to the manufacturing technique used to prepare the specimens, and can be efficiently suppressed via thermal annealing. In such samples that were subjected to a high-temperature annealing before laser exposure, a stable stress level is observed for stresses below as well as above 1 GPa.

Finally, our unique combination of flexure-based tensile tester micro-device operated with a laser with photoelasticity measurements offers an efficient method to explore the mechanical behavior of glass at time-and-length scales that could not be accessed before experimentally [10], [11], [38], [39], and that too, under various experimental conditions.

## 7  Acknowledgments and author contributions

In this work, S.I.N designed the tensile tester device, constructed an improved experimental setup inspired from a previous version realized by C.E.A, performed the measurements, and wrote the draft manuscript. S.I.N and Y.B discussed and interpreted the results. C.E.A performed preliminary and preparatory experiments on this topic. Y.B proposed the initial concept, designed, and supervised the research. All authors contributed to further revisions of the manuscript.

The Galatea Lab acknowledges the sponsoring of Richemont International. The funding support of the European Research Council is also acknowledged (ERC-2012-StG-307442 and ERC-2017-POC-790169).